\documentclass[prl,twocolumn,groupedaddress,showpacs]{revtex4}
\usepackage{graphicx}
\usepackage{latexsym}
\usepackage{amssymb}
\usepackage{amsfonts}
\begin{document}
\bibliographystyle{apsrev}
\title{Vortex-antivortex dynamics and field polarity dependent flux creep in hybrid superconductor/ferromagnet
nanostructures}
\author{M. Lange}
\altaffiliation{Present address: nv Bekaert sa, Bekaertstraat 2, 8550 Zwevegem,
Belgium}
\author{M. J. Van Bael}
\author{A. V. Silhanek}
\author{V. V. Moshchalkov}
\email{victor.moshchalkov@fys.kuleuven.ac.be}
\affiliation{Nanoscale Superconductivity and Magnetism Group, Laboratory for Solid
State Physics and Magnetism, K.U.Leuven, Celestijnenlaan 200D, 3001 Leuven, Belgium}
\date{\today}
\begin{abstract}
Vortex-antivortex arrays (VAA) that are created in a type-II superconducting film by
lattices of ferromagnetic dots with perpendicular anisotropy have been investigated.
The highest critical current is shifted to a non-zero value of the applied field, and
current - voltage characteristics show different regimes of vortex motion even in zero
applied field due to the presence of the VAA. Creep of interstitial vortices is
observed at low driving forces. This flux creep is strongly field polarity dependent.
\end{abstract}
\maketitle
\begin{figure}
 \includegraphics{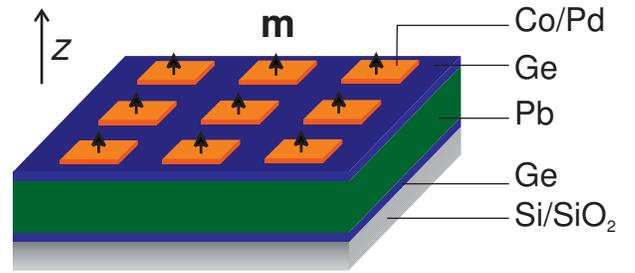}
\caption{Schematic drawing of the investigated superconducting / ferromagnetic hybrid
structure. The superconductor is a 85~nm Pb film, which is covered by an array of
ferromagnetic Co/Pd dots with out-of-plane magnetization.}
 \label{schema}
\end{figure}

\begin{figure}
 \includegraphics{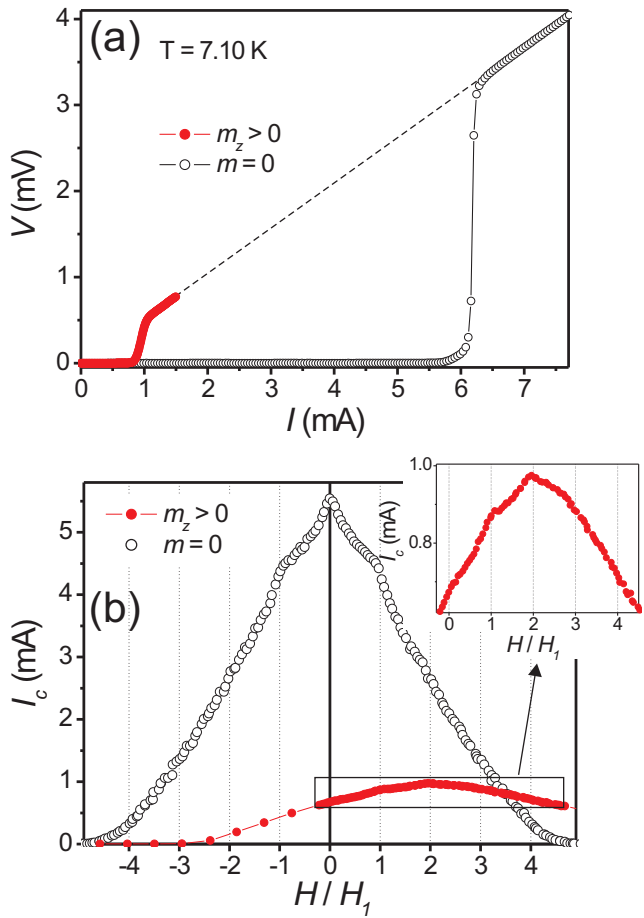}
\caption{(a)~$V(I)$ curves at $T=7.10$~K in zero field after demagnetization ($m=0$)
and after magnetization of the sample in a high positive field ($m_z>0$). (b)~Field
dependence of the critical current of the superconductor with the dots in the two
different magnetic states. A voltage criterion of $V_{cr} = 1~\mu$V was used.}
 \label{iccom}
\end{figure}

\begin{figure}
 \includegraphics{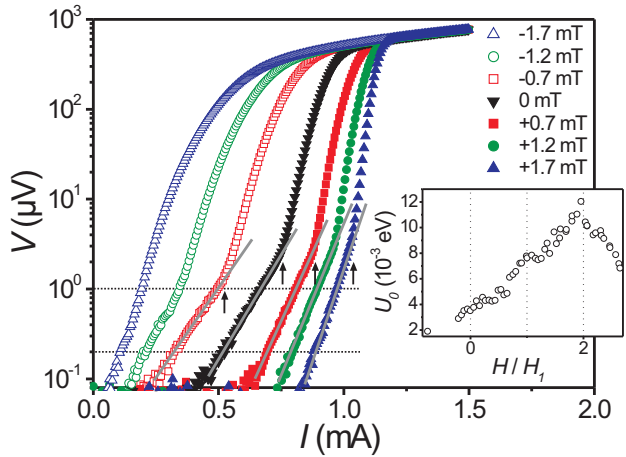}
\caption{Field polarity dependent flux creep is observed in
current - voltage characteristics of the superconductor at
$T=7.10$~K. The dots are in the $m_z>0$ state and $V$ is plotted
in logarithmic scale. The full lines are guides to the eye. The
arrows indicate the crossover between low and high current
regimes.  The dotted lines indicate the used voltage criteria for
determination of $I_c(H)$ (see text). The inset shows the field
dependence of the activation energy $U_0$ determined from the
slope of the $\ln V (I)$ curves.}
 \label{ivcreep}
\end{figure}

\begin{figure}
 \includegraphics{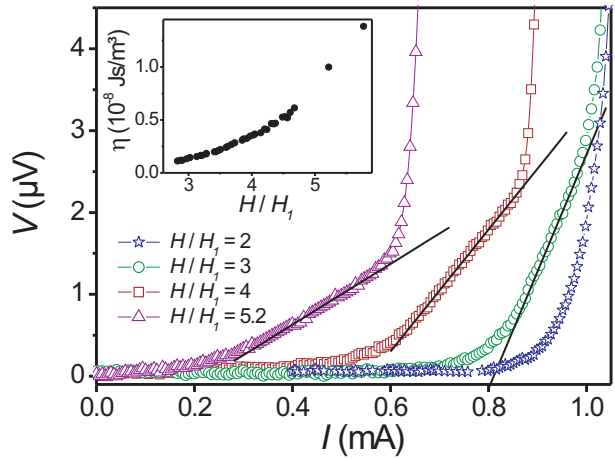}
\caption{Current - voltage characteristics of the superconductor at $T=7.10$~K in the
$m_z>0$ state at $H/H_1 = 2$, 3, 4 and 5.2 for small voltages. The slope of the
straight lines determines the viscous drag coefficient $\eta$ in the flux flow regime
$H > 2.8$.}
 \label{ivflow}
\end{figure}

 In the mixed state magnetic field penetrates type-II
superconductors in the form of vortices carrying a quantized
magnetic flux, which normally corresponds to a single flux quantum
\cite{abrikosov57}. In order to avoid dissipation in
superconductors it is necessary to prevent vortex motion.
Effective pinning can for instance be achieved by introducing
columnar defects \cite{civale91,thompson97}, lattices of submicron
holes (antidots) \cite{hebard77,baert95a,metlushko99,silva01} or
ferromagnetic dots \cite{geoffroy93,martin97,morgan98,vanbael99}.
Besides enhanced critical currents, {\em field polarity dependent}
flux pinning was observed by Morgan and Ketterson in
superconducting Nb films with arrays of ferromagnetic Ni dots
\cite{morgan98}, whose magnetization has an out-of-plane
component. It was shown that in this system the critical current
density $j_c$ depends on the mutual orientation of applied field
and magnetization of the dots. Stable vortex-antivortex arrays
(VAA) are predicted to exist in hybrid
superconductor/ferromagnetic systems even in zero external field
due to the inhomogeneous stray fields of the ferromagnets
\cite{marmorkos96,milosevic02,milosevic03,milosevic04,besp01a,erdin02a,
erdin02b,aladyshkin03,laiho03,priourla,carneirola}. Despite strong
interest of theory in this subject, experimental reports on VAA
have been very rare \cite{vanbael01}. Experiments on magnetic dots
with out-of-plane magnetization have been carried out only for
relatively weak stray fields in the sense that the stray field was
screened by Meissner currents in the superconductor
\cite{vanbael00,vanbael03,grigorenko03}.\\ In this paper we will
show that the vortex dynamics in type-II superconducting Pb films
is strongly influenced by the presence of ferromagnetic Co/Pd dots
with perpendicular anisotropy. We will characterize the dynamic
behavior of stray field-induced VAA by means of current ($I$) -
voltage ($V$) characteristics and demonstrate pronounced field
polarity dependence of the flux creep. A clear sign for the
existence of VAA is the presence of
different regimes of vortex motion even in zero applied field.\\
Fig.~\ref{schema} shows a schematic drawing of the investigated
sample. The structural and magnetic properties of this sample have
been described elsewhere \cite{lange03}.
The sample consists of a type-II superconducting Pb film
(thickness $t_{Pb} = 85$~nm, critical temperature $T_c = 7.26$~K)
evaporated on a 1~nm Ge base layer on a Si substrate with
amorphous SiO$_{2}$ top layer. The Pb film is covered by a 10~nm
Ge layer, which prevents the influence of the proximity effect
with the metallic dot array. In order to avoid inhomogeneities of
the current, this continuous Ge/Pb/Ge trilayer is patterned into a
transport bridge (width $w = 200$~$\mu$m, distance between voltage
contacts $d = 630$~$\mu$m) by optical lithography and chemical wet
etching. Measurements of the upper critical field of this
transport bridge allow an estimation of the coherence length
$\xi(0)= 34$~nm and of the penetration depth $\lambda(0) = 49$~nm.
The transport bridge is covered by ferromagnetic dots using
electron-beam evaporation and electron-beam lithography. The dots
consist of a 3.5~nm Pd base layer and a
[Co(0.4~nm)/Pd(1.4~nm)]$_{10}$ multilayer with perpendicular
magnetic anisotropy \cite{carcia85}. The dots are arranged in a
square array with period $L = 1.5$~$\mu$m. They have a square
shape with a side length of approximately 0.8~$\mu$m with slightly
irregular edges.\\ The vortex pinning properties of this hybrid
sample were investigated by electrical transport measurements for
two different magnetic domain states of the Co/Pd dots. After
aligning the magnetic moments of the dots in positive
$z$-direction ($m_z>0$) (by magnetizing the dots in a high
positive applied field of $\mu_0 H = +1$~T perpendicular to the
sample surface) the dots are in a {\em single domain state}
\cite{lange03}. After an ac demagnetization, the dots are in a
{\em multidomain state} where the net magnetic moment of each dot
$m$ is zero ($m=0$) \cite{lange03}. Note that the fields applied
during all measurement are much weaker than the coercive field of
the dot array at room temperature $\mu_{0} H_{coe} = 150$~mT.
\\ $V(I)$ curves of the sample are measured in a Quantum Design
Physical Properties Measurement System (PPMS) in standard
four-point geometry by increasing $I$ starting from $I=0$.
Figure~\ref{iccom}(a) shows two $V(I)$ curves measured with the
sample in the $m=0$ and the $m_z>0$ state at temperature
$T=7.10$~K in zero field.
There is a remarkable difference between the two curves: In the
magnetized state $V$ shows a steep transition to the normal state
at about 6~mA, while in the $m_z>0$ state this transition already
appears around 1~mA.\\ We define the critical current $I_c$ as the
current where the voltage exceeds a certain voltage criterion
$V_{cr}$. The critical current density $j_c$ can be calculated
using $j_c = I_c / (t_{Pb} \, \, w)$. The field dependence of
$I_c$ at $T=7.10$~K is shown for $V_{cr} = 1~\mu$V in
Fig.~\ref{iccom}(b) for both magnetic states. The first matching
field $H_1$ of this sample corresponds to the field value which
generates one flux quantum per unit cell of the dot array and is
given by $\mu_0 H_1 = \phi_0 / L^2 = 0.92$~mT ($\phi_0$ is the
superconducting flux quantum). In the $m=0$ state the maximum of
$I_c$ is observed at zero field, and the $I_c(H)$ curve is
symmetric with respect to $H=0$. Clear matching effects appear at
positive and negative first matching field, indicating that the
multidomain dots act as pinning sites for vortices due to the
local suppression of the order parameter under the dots caused by
the stray field, or due
to the magnetic pinning of the vortices by the domain structure.\\
In the $m_z>0$ state $I_c$ is lower than in the $m=0$ state for
$H/H_1<3.5$. Surprisingly, the highest value of $I_c$ is not
observed at $H=0$, but at $H/H_1 = 2$. Clear matching features are
seen at $H/H_1 = 0$, 1 and 2.  The strong overall reduction of the
critical current indicates that the shift is not due to a mere
compensation effect of the applied field by the local stray field
of the dots \cite{lange03}, but that the vortex dynamics is
strongly modified due to vortices induced by the dots, as will be
explained below. \\
The significant alteration of $I_c(H)$ which is obtained only by a
different magnetic history of the sample must be related to the
changed magnetic domain state of the dots. It should be noted that
$H$ denotes only the externally applied field.  The local stray
field of the dots is also present at $H = 0$.  However, when
integrated over the superconducting layer, the net average stray
field of the dots will be near zero because of the returning stray
field of opposite sign. For the further analysis of the data we
neglect the influence of heating effects, nonuniform current
distributions and surface barriers since they are the same in both
magnetic states and cannot account for the observed significant
change in $I_c(H)$. In the following we will show that the
existence of {\em two vortices or one giant vortex with vorticity
two} directly beneath each dot can consistently explain the
anomalous field-shift and also other features in the $V(I)$ curves
that will be discussed below. Additionally the returning stray
field of the dots generates {\em two antivortices} per unit cell
of the dot array in the superconductor. These antivortices are
located at interstitial positions of the dot array. This
assumption is justified by several theoretical calculations that
predict this picture as a possible scenario
\cite{milosevic02,milosevic03,milosevic04,priourla}. In the case
of an isolated dot both, vortices and antivortices, are bound to
the dot, but for a regular array of dots the periodically
modulated stray field could change this picture. If the spacing
between the dots is sufficiently small, a motion of vortices and
antivortices should be possible when a transport current exerts a
sufficiently strong driving force on them.  \\
Fig.~\ref{ivcreep} shows the $V(I)$ curves measured at $H = 0$,
$\pm 0.7$~mT, $\pm 1.2$~mT and $\pm 1.7$~mT with $V$ plotted on
logarithmic scale.
Low and high current regimes must be distinguished: In the low
current regime only interstitial flux lines move, whereas at
higher applied current both interstitial flux lines and the ones
that were originally pinned on the dot sites move. Different
dynamic regimes for low and high currents have also been observed
in $V(I)$ curves of superconducting films with arrays of antidots
\cite{rosseel96} and arrays of magnetic dots \cite{villegas03}.
Thus, the $V(I)$ curve at $H=0$ is in agreement with the
assumption that interstitial antivortices are present even in zero
field.\\ In the low current region ($V \alt 5$~$\mu$V) the $\ln V
(I)$ curves measured in zero field and positive fields have a
linear slope as is indicated by the straight lines. This linear
slope is less clear for the curve at $-0.7$~mT and disappears as
$H$ is further decreased. A linear $\ln V(I)$ dependence is
predicted in the Anderson and Kim model for flux creep
\cite{anderson64}. In this model the electric field $E$ generated
by the vortices creeping from one pinning site to another in the
limit $I \approx I_c$ is given by
\begin{eqnarray}
E \propto e^{\frac{U_0}{k_B T}\left(\frac{I}{I_c}-1\right)} \qquad I \approx I_c
\label{fcreep},
\end{eqnarray}
with $k_B$ the Boltzmann constant and $U_0 = U_0(T,H)$ the activation energy, i.e. the
height of the potential barrier between two adjacent pinning sites. Thus,
\begin{eqnarray}
\ln V = \frac{U_0}{k_B T}(\frac{I}{I_c}-1) + C(H,T),
\end{eqnarray}
where C is independent of $I$. This means that in zero applied
field the experimentally observed $\ln(V) \propto I$ behavior is
consistent with a creep of antivortices over the potential
barriers that are formed by the dots and the induced vortices.
{\em Hence, Fig.~\ref{ivcreep} shows that flux creep in hybrid
superconductor/ferromagnet systems is strongly depending on the
field polarity.} The exact creep pattern of the antivortices in
the sample remains an open question. An antivortex can for example
move by anihilating first with a vortex on the edge of a dot, and
at the same time a new antivortex is generated at the opposite dot
edge. Another possibility is that the antivortices creep through
channels between the dots. A theoretical treatment in the
framework of the time-dependent Ginzburg-Landau theory is required
to answer this question.\\ The slope of the $\ln V(I)$ curves is
only determined by $U_0$, $T$ and $I_c$. For the calculation of
$U_0$ (results are shown in the inset of Fig.~\ref{ivcreep}) we
have defined $I_c$ with a criterion of $V_{cr} = 0.2$~$\mu$V. Up
to $H_2$, both $U_0$ and $I_c$ increase with $H$, which is related
to the decreasing number of antivortices. As the number of
antivortices goes down, the effective barrier that they have to
overcome is increased due to a decreasing mutual repulsive
interaction. Note that this interaction is not included in the
model (in its simplest interpretation). At $H_1$ a peak in
$U_0(H)$ is observed, because of the commensurability of the
antivortices with the periodic pinning potential. Above $H_2$,
$U_0$ starts to decrease drastically as $H$ is increased. This
crossover can be explained by the appearance of vortices at
interstitial positions of the dot array: the number of vortices
increases and therefore $U_0$ decreases with increasing $H$. The
most important conclusion of these results is that {\em the
observed features are in good agreement with the assumption that
at $H=0$ two antivortices per unit cell are induced in the
superconductor at interstitial positions of the dot array}.  Small
inhomogeneities of the dots in a real sample could lead to a
variation of the stray fields of the dots, which for a small
fraction of dots could result in three or only one induced
vortex-antivortex pair instead of two. However, this will not
influence the general behavior of
the sample.  \\
A crossover to a linear $V(I)$ dependence is observed for
$H/H_1>2.8$. The main panel in Fig.~\ref{ivflow} shows the low
voltage part of the $V(I)$ curves in the $m_z>0$ state for $H/H_1
= 2$, $3$, $4$ and $5.2$.
The linear dependence is consistent with the situation that local
pinning forces are overcome resulting in a steady motion of the
interstitial vortices \cite{silhanek03} with velocity $v$ given by
\begin{eqnarray}
\eta v = (I - I_c) \phi_0 / (t_{Pb} \, \, w) \label{flow},
\end{eqnarray}
with $\eta$ a viscous drag coefficient, which can be determined from
\begin{eqnarray}
V = E \, d = n_f \, \phi_0 \, v \, d = n_f \phi_0^2 d (I-I_{c,flow})/(\eta \, \,
t_{Pb} \, \, w) \label{flow2},
\end{eqnarray}
where $n_f$ is the areal density of moving interstitial vortices
given by $n_f = \mu_0 |H - H_2|/\phi_0$. $I_{c,flow}$ is defined
as the value of $I$ where the straight lines characterizing the
flux flow regime indicated in Fig.~\ref{ivflow} cross the
$I$-axis. From the slope of the $V(I)$ curves and the value of
$I_{c,flow}$, $\eta(H)$ is calculated using eq.~(\ref{flow2}), see
the inset of Fig.~\ref{ivflow}. In the Bardeen-Stephen model for
free flux flow \cite{bardeen66} it is assumed that $\eta$ does not
depend on the applied field. The flux flow phase of the
interstitial vortices in our measurements can therefore not be
interpreted in terms of free flux flow. The mutual interaction
between flowing interstitial vortices and pinned vortices on the
dot sites can not be neglected and should be the main reason for
the increase of $\eta$ as $H$ is increased.\\ Finally, we would
like to point out that ferromagnetic dots can be very useful for
the new field of {\em fluxonics}. Here the aim is to guide the
vortices through nanoengineered arrays of pinning sites or
channels in the superconductor \cite{silhanek03}, which makes it
possible to develop new devices, like superconducting reversible
rectifiers \cite{villegas03} or ratchet cellular automata based on
nanostructured superconductors \cite{hastings03}. Using VAA in a
well-designed fashion makes it possible to operate fluxonic
devices without exposing them to externally applied fields. For
instance, by using triangular shaped magnetic dots it is possible
to measure DC voltages upon application of an AC current
\cite{villegas03} even in zero field. Furthermore, by using
rectangular dots that generate one vortex-antivortex pair per dot,
logic elements can be developed \cite{hastings03} where the 0 and
1 state are defined by the position of the antivortex.\\ In
conclusion, vortices in hybrid superconductor/ferromagnet
nanostructures can have peculiar dynamic properties due to the
presence of the vortex-antivortex arrays that are induced in the
superconductor by the inhomogeneous stray field of the
ferromagnet. In zero applied field two regimes of vortex motion
are observed, which proofs the existence of these induced
vortex-antivortex arrays. For the first time field polarity
dependent flux creep is demonstrated.\\ The authors are thankful
to E. Claessens for help with the measurements, to S. Raedts, M.
Morelle and K. Temst for their contribution to the sample
preparation, and to Y. Bruynseraede for fruitful discussions. This
work was supported by the Belgian IUAP, the ESF "VORTEX" and the
K.U.Leuven Research Fund GOA/2004/2 programs, and by the Fund for
Scientific Research - Flanders (Belgium) (F.W.O.-Vlaanderen).
M.J.V.B. is a Postdoctoral Research Fellow of the
F.W.O.-Vlaanderen.
\newcommand{\noopsort}[1]{} \newcommand{\printfirst}[2]{#1}
  \newcommand{\singleletter}[1]{#1} \newcommand{\switchargs}[2]{#2#1}
  \newcommand{\bibi}[1]{\bibitem{{#1}}{\small{(#1})}}

\end{document}